# Ultrafast molecular dynamics of dissociative ionization in OCS probed by soft X-ray synchrotron radiation

**Ali Ramadhan[1], Benji Wales[1], Isabelle Gauthier[2], Reza Karimi[3], Michael MacDonald[2], Lucia Zuin[2], Joe Sanderson[1]**

[1]Department of Physics and Astronomy, University of Waterloo, 200 University Avenue West, Waterloo, Ontario, Canada N2L 3G1
[2]Canadian Light Source, 44 Innovation Boulevard, Saskatoon, Saskatchewan, Canada S7N 2V3
[3]Department of Energy Engineering and Physics, Amirkabir University of Technology, 424 Hafez Avenue Tehran, Iran 15875-4413

Email: j3sanderson@uwaterloo.ca

**Abstract**
Soft X-rays (90-173 eV) from the 3$^{rd}$ generation Canadian Light Source have been used in conjunction with a multi coincidence time and position sensitive detection apparatus to observe the dissociative ionization of OCS. By varying the X-ray energy we can compare dynamics from direct and Auger ionization processes, and access ionization channels which result in two or three body breakup, from 2+ to 4+ ionization states. We make several new observations for the 3+ state such as kinetic energy release limited by photon energy, and using Dalitz plots we can see evidence of timescale effects between the direct and Auger ionization process for the first time. Finally, using Dalitz plots for OCS$^{4+}$ we observe for the first time that breakup involving an O$^{2+}$ ion can only proceed from out of equilibrium nuclear arrangement for S(2p) Auger ionization.

## 1. Introduction

Coulomb Explosion Imaging (CEI) is a developing technique designed to image the geometry of small molecules (<10 atoms) as they bend, stretch, and separate on a femtosecond time scale ($10^{-15}$ s). The technique usually relies on ultrafast infrared laser pulses [1,2] or highly charged ion impacts [3] but X-ray and UV laser pulses from free-electron laser sources have also been used [4]. Here we use CEI utilising ionization by single X-ray photons from a third generation synchrotron source.

The (near) simultaneous ejection of more than one electron from an atom in response to absorption of an X-ray photon is the result of electron-electron correlations and as a result has been extensively studied in helium and other atoms [5]. At energies away from threshold this is dominated by two mechanisms. The first TS1, or "knock-out", is where the outgoing photoelectron scatters of the other electrons to produce multiple ionization [6]. The second, shake-off [7], is where the rapidly outgoing photoelectron leaves the atom in a suddenly different charge state. In this case there is a possibility that an electron experiencing this sudden difference in effective charge may find itself in the continuum. The end result is that the branching ratio of double to single ionization in atoms generally rises from zero at threshold to a plateau where the total excess energy is around 100 eV (e.g for helium [8]) and that the energy sharing between the two electrons involved is generally unequal [9]. Hartman et al. [10,11] have extended these mechanisms from atoms to moderately large molecules. We assume that these mechanisms are available for the double and triple ionization of OCS at 90 eV.

As well as direct ionization the processes above may leave an intermediate ion in a highly excited state, which can then shed a further electron through autoionization. Depending on the lifetime(s) involved this may occur at any point in the dissociation process up to and including a fragment atom. For a diatomic molecule this can be described as $AB + h\nu \rightarrow (AB^+)^* \rightarrow A^+ + B^* \rightarrow A^+ + B^+$. Price and Eland [12] describe a particularly nice example for $O_2$. Extensions to a triatomic molecule are easily imagined. This may produce a low energy electron.

Above an energy where a core electron can be excited another mechanism for double and multiple ionization becomes available: Auger ionization. Here a core electron is either excited into a vacant orbital (resonant Auger) or into the continuum (normal Auger). Subsequently the core hole is filled by an electron in a higher energy level, with the excess energy being carried away by the ejection of a third electron. This process has a distinct timescale and typically produces electrons of a distinct energy. During this Auger lifetime the nuclei of the molecule may move, as determined by the potential energy surfaces of the excited neutral (resonant Auger) or excited ion (normal Auger). Of course the Auger electron may interact with the remaining electrons in the atom or molecule through the TS1 or shake-off mechanism described above. Excitation at 167 eV gives access to resonant Auger processes, while excitation at 172 eV and 173 eV gives access to normal Auger processes.

After a triatomic molecule absorbs a photon in the soft X-ray wavelength range, it ionizes and loses up to 4 electrons through direct or Auger ionization. It can then dissociate into two or three fragments, two or more of which are ions. In the case when the molecule decays into its constituent atomic ions, the break up can be modeled as a decay through the Coulomb repulsion, in which an imprint of the initial molecular shape is left on the final fragment momenta. The extent to which this is an accurate description depends in part on the speed of a multiple ionization event, by which we refer to the time between the initial and final ionization steps, during which the charge builds up. The removal of the initial electron in both direct and Auger processes is on an attosecond timescale. If the multiple ionization is rapid compared to nuclear motion (t < 10 fs) then the initial molecular geometry may be preserved, but if the event is slow, (10 < t < 100 fs), and the initial geometry is known, then nuclear dynamics during the ionization process can be revealed in the pattern of momentum release [2].

We attempt to vary the ionization timescale in this work by scanning the photon energy through the sulfur 2p edge at ~172 eV. Below this edge, ionization is dominated by direct ionization in which electrons are removed promptly by photon absorption, while above the edge the ionization is dominated by electron ejection due to Auger decay which through its lifetime (t ~ 10-100 fs), extends the duration of the ionization event.

## 1.1 Valence absorption and ionization

The photoabsorption, photoionization, and photofragmentation of OCS has been well studied. The ultraviolet absorption cross section has been measured [13] as has the valence photoionization cross section [14], threshold photoelectron spectrum [15], He(i) photoelectron spectrum [16–18], and electron energy loss spectrum [19]. The inner valence region has been probed by photoabsorption [20], electron energy loss [21], photoelectron spectroscopy using synchrotron [22,23], He(ii) [24] and Mg K-α [25] radiation, and studied theoretically [26,27]. Autoionization from S atoms has been observed from photoexcited OCS [28].

## 1.2 Inner valence ionization and double ionization

Coincidence spectroscopies allow higher electronic states and higher charge states to be probed. The dication of OCS has been probed by TEPEsCO [29] with the threshold found at 30.0±0.1 eV. Various electron multi-ion time-of-flight spectroscopies have been used to study the fragmentation of valence [30–32] and inner valence [33–35] excited OCS. We note that Masuoka [31] gives the threshold for double photoionization as 32 eV and the threshold for triple ionization as 52.5 eV. The maximum in the triple ionization cross section (below any S(2p) absorption threshold) is reported as 95 eV. Excitation of OCS at 90 eV can result in the direct production of a triply ionized molecule through the "knock out" process, without the effects of a S(2p) core-hole lifetime. Masuoka also reports the branching ratios for all ion production up to 100 eV photon energy [30]. Eland et al have reported double [36] and triple [37] coincidences between $O^+$, $C^+$ and $S^+$ from ionization of OCS at 65 eV with the $S^+$ and $O^+$ fragments in the triple ionization taking away more initial kinetic energy than in the double ionization case.

## 1.3 Sulfur 2p edge ionization

Absorption spectra and total ion yield spectra have been recorded in the region of the Sulfur 2p edge [38–41]. Theoretical multi-reference, single and double-excitation configuration interaction, ab initio (MRD-CI) calculations have lent weight to the peak assignments [42]. Using dipole (e, e plus ion) coincidence spectroscopy, Feng et al. [35] have measured branching ratios for the production of fragment ions at equivalent photon energies from the first ionization threshold up to 300 eV. Ion fragmentation studies have been carried out using the time-of flight coincidence method near the sulfur 2p edge. Erman et al. [40] have measured the photofragment ion asymmetry parameter (beta) across the sulfur 2p absorption edge, as well as the PePIPICO maps for double ionization events, which have also been measured by Franzen et al. [43]. Ankerhold et al. [39] have measured the total charge production in OCS as one scans across the Sulfur 2p edge. In the same paper they report the maximum kinetic energy of the fragments from both double and triple photoionization. As expected, the fragments from triple ionization events take away considerably more energy than those from double ionization events. Another class of electron ion measurements is Molecular Frame Photoelectron Angular Distributions (MFPADs). Here the photoelectron is measured in coincidence with an ionic fragment from the parent molecule to give a "fixed in space" angular distribution of photoelectrons. This has been reported for OCS by Golovin et al. [44] for OCS at the $S2p_{3/2}$, $S2p_{1/2}$, C1s, and O1s electrons.

Sulfur 2p photoelectron and Auger electron processes have been well studied. Carrol et al. [45] have recorded Auger spectra for C(1s) O(1s) and S(2p) ionization and determined level assignments for the final states of the 2+ ion. Bolognesi et al. [46] have measured the S L23MM Auger electron spectrum in coincidence with the photoelectron, since the energy range of the Auger electrons is 127 eV to 141 eV this Double Ionization Energy (DIE) ranges from 30 eV (threshold) to 45eV. Niskanen et al. [47] have measured the energies of core valence doubly excited OCS though double ionization photoelectron spectroscopy (DIPES) which is essentially a variant on TOF-PEPECO. This allows them to directly measure OCS S $2p^{-1}v^{-1}$ double ionization energies. These start at approximately 190 eV and are above the photon energies used in this study. Saha et al. [48] have measured the fragmentation of OCS in coincidence with the Auger electron. Their experimental set-up – a Cylindrical Mirror analyser (CMA) for the electron detection and an ion time-of flight system with position sensitive detection for the ions – allowed them to measure the mean kinetic energy release (KER) upon fragmentation of $OCS^{2+}$ for different Auger states. The fragmentation of the molecule was found to be highly dependent on different Auger decays through which the precursor molecular ions are formed. Carlson et al. [49] have reported that following resonant absorption the excited electron acts as a spectator electron in the subsequent Auger decay.

1.4 Higher energy ionization
Of the papers cited above only the dipole coincidence spectroscopy papers of Feng et al. [21,35] cover the sulfur 2s edge. The photofragmentation studies and ion coincidence studies above, likewise, do not cover the S(2s) absorption edge.

Hikosaka et al. [50] have studied the Sulfur (2s) Auger spectrum at sub natural linewidth resolution. This was done by exciting the OCS S(2s) photoelectron at threshold and measuring Auger threshold coincidences. Exciting the S(2s) electron differs from exciting the S(2p) toelectron in that the major Auger component is the Coster-Kronig decay of the 2p electron to fill the 2s hole, producing an Auger electron of energy 45 eV or lower. They find that the core-hole lifetime of the 2s hole corresponds to a width of 1.8 eV (0.37 fs). By contrast the S 2p core-hole lifetime is 65 meV (10.1 fs) [51] in OCS. Following the Coster-Kronig decay the 2p hole may be filled by further Auger transitions leading to higher charge states of OCS and its fragments being detected.

Sham et al. [52] present a total ion yield spectrum for OCS across the C(1s) edge, as does Erman et al. [40]. The PePIPICO studies of Franzen [43] and Erman [40] also cover the C(1s) edge with Erman [40] also including a total electron yield spectrum across the C(1s) edge. Using a momentum imaging time of flight spectrometer, Laksman et al. [53] have measured the anisotropy parameter (beta) for OCS dissociated to $O^+ + CS^+$ and $S^+ + CO^+$. The dissociation necessarily producing two ions at 180° separation. They found that the intensity for the $S^+ + CO^+$ channel was always 10x to 15x greater than the $O^+ + CS^+$ channel. They also found that the anisotropy of the $O^+ + CS^+$ channel was greater than that of the $S^+ + CO^+$ channel, which they suggest, implies the dication state leading to $S^+ + CO^+$ has significant bending vibrational excitation. The paper also describes the total kinetic energy release and momentum correlations for OCS producing three ions $O^+ + C^+ + S^+$ on and off resonance at the C(1s) edge, with the total kinetic energy release ranging from 10 to 30 eV both on the 1s-π* and 1s-σ* resonance and off resonance. They investigated bending of OCS for the C(1s) excitation and O(1s) excitation and observe effects of Renner-Teller splitting for C(1s) excitation only, which were distinguishable by their respective molecular orientations relative to the photon polarization.

1.5 Ionization at the S(1s) edge
Ankerhold et al. [54] have measured the charged particle production in coincidence for photoexcitation of OCS both above and below the S(1s) edge. They find the most abundant combination excited below the edge (2449 eV) to be $O^+ + C^+ + S^+$ and above the edge to be $O^+ + C^+ + S^{2+}$. They also plot the relative abundances of the various (total charge = 2 and 3) channels across the S(1s) edge. Neville et al. [55] have recorded a total ion yield spectrum across the S(1s) edge with traces for various individual ion channels as well as PePIPICO maps for double ion and triple ion production. Bomme et al. [56] present a total ion yield spectrum for OCS excited near the S(1s) edge, complete 2D time of flight maps for the production of $CO^+ + S^+$, $CO^+ + S^{2+}$, $CO^+ + S^{3+}$, and $CO^{2+} + S^{2+}$ dissociation channels. They also present complete 3D MFPADs for dissociation to $S^+ + CO^+$, 10 eV above threshold.

1.6 Excitation by highly charged ions and ultrafast laser pulses

Recent electron capture work employing highly charged ion impact experiments with $Ar^{4+}$ and $Ar^{8+}$ projectiles at 15 keV $q^{-1}$ [3,57,58] has also probed the break up of OCS from 2+ up to 6+. Findings showed energy release from the 3+ state to be lower than the calculated Coulombic value, in agreement with recent femtosecond laser induced ionization work [59]. Higher charge states peaked closer to the coulombic value but with distributions wider than the laser work results, indicating the presence of more excited states in the HCI impact induced ionization than the tunnel ionization process associated with the femtosecond laser induced ionization.

## 2. Experimental

The experiment was performed at the Variable Line Spacing Plane Grating Monochromator (VLS-PGM) soft X-ray beam line at the 3rd generation Canadian Light Source (CLS) in Saskatoon, Saskatchewan, Canada [60]. A portable Coulomb Explosion Imaging apparatus from the University of Waterloo was installed at the beamline behind a differential pumping chamber. Adapted from the single stage time of flight (TOF) apparatus of [61], an effusive gas beam of OCS is injected into a high-vacuum chamber through a hypodermic needle placed one millimeter away from an orthogonal 1mm diameter, collimated beam, of variable energy X-ray photons (5.5 to 250 eV) with a resolution of >10,000 E/ΔE. The beam enters the chamber horizontally and the electric vector of the 100% linearly polarized beam is horizontal (perpendicular to the TOF axis). The photon flux was reduced using the entrance and exit slits and baffles such that a typical count rate was of the order of 100 per second. Electrons produced via direct ionization and Auger processes in the interaction region are accelerated downward with a constant electric field (26,710 V/m) through a 2.9 mm aperture and are detected with a microchannel plate (MCP), operating as the time of flight start signal. The aperture and field are such that electrons with a kinetic energy component orthogonal to the electric field of more than 1 eV will not pass through the aperture. Thus the aperture becomes an increasingly efficient filter for electrons with a total kinetic energy above 1 eV and our apparatus biases in favour of low energy electrons. This energy filtering biases our detector in favour of 1) low energy photoelectrons produced by the photoionization of S(2p) core states at 172 eV and 173 eV and 2) low energy electrons produced through the TS1 and shake-off mechanisms described in the Results section. This biases us against straightforward single ionization from valence states [31]. The electron aperture is aligned directly below the interaction region in order to filter out interactions with the background gas, allowing us to define a zero momentum position for all fragments, which is crucial in calculating fragment momenta. Positive ion fragments on the other hand are accelerated upwards. The fragments are detected and their full 3D momenta are determined using a time and position-sensitive capacitively coupled film anode detector at the end of the spectrometer [60,62]. The four outputs from the anode are recorded by a pair of Gage oscilloscope cards interfaced with a PC giving a maximum recording rate of 1 kHz. The atomic and molecular fragment ions were measured in double or triple coincidence: $CO^{a+} + S^{b+}$, $O^{a+} + CS^{b+}$, or $O^{p+} + C^{q+} + S^{r+}$ ($a + b$ = 2 or 3, $p + q + r$ = 3 or 4). Ions are identified as the result of the fragmentation of a single molecule only if their total momentum is close to zero (< $5\times10^{-23}$ kg m/s).

Although the asynchronous nature of the data acquisition means that the probability of detecting ions from more than one molecule for any electron start is small, data analysis which can distinguish between all possible combinations of ions given by their arrival times is used to identify true coincidences [63].

## 3. Results

Data was collected at 90, 166.94, 171.6, and 172.6 eV. 90 eV was chosen to observe the ground state geometry and locally maximizes ionization efficiency [31]. 166.94 eV was chosen as it was below the S(2p) edge and coincided with the $S(2p_{3/2})^{-1} \rightarrow$ 4s Rydberg state transition [40]. 171.6 eV was chosen as it ionized the $S(2p_{3/2})$ electron to the continuum, while 172.6 eV was chosen as it ionized both the $S(2p_{3/2})$ and $S(2p_{1/2})$ electrons to the continuum [64].

Throughout this section, photon energy figures are rounded to the nearest eV for brevity, thus the 90, 166.94, 171.6, and 172.6 eV energies are referred to as 90, 167, 172, and 173 eV respectively. The ($a,b,c$) notation refers to the $OCS^{q+} \rightarrow O^{a+} + C^{b+} + S^{c+}$ ($q=a+b+c$) fragmentation channel. For three body breakup, we observe and discuss the (0,1,1), (1,0,1), (1,1,0), (1,1,1), (1,1,2), (1,2,1), and (2,1,1) channels. For two body breakup, the $O^+ + CS^+$, $CO^+ + S^+$, and $CO^+ + S^{2+}$ channels are observed and discussed. All other possible fragmentation channels were too rare to offer enough statistics for discussion.

3.1 2-ion kinetic energy release

Kinetic energy release (KER) distributions for various double coincidence (2-ion) fragmentation channels are plotted in figure 1. The kinetic energy of each fragment ion is calculated from its mass and experimentally measured momenta.

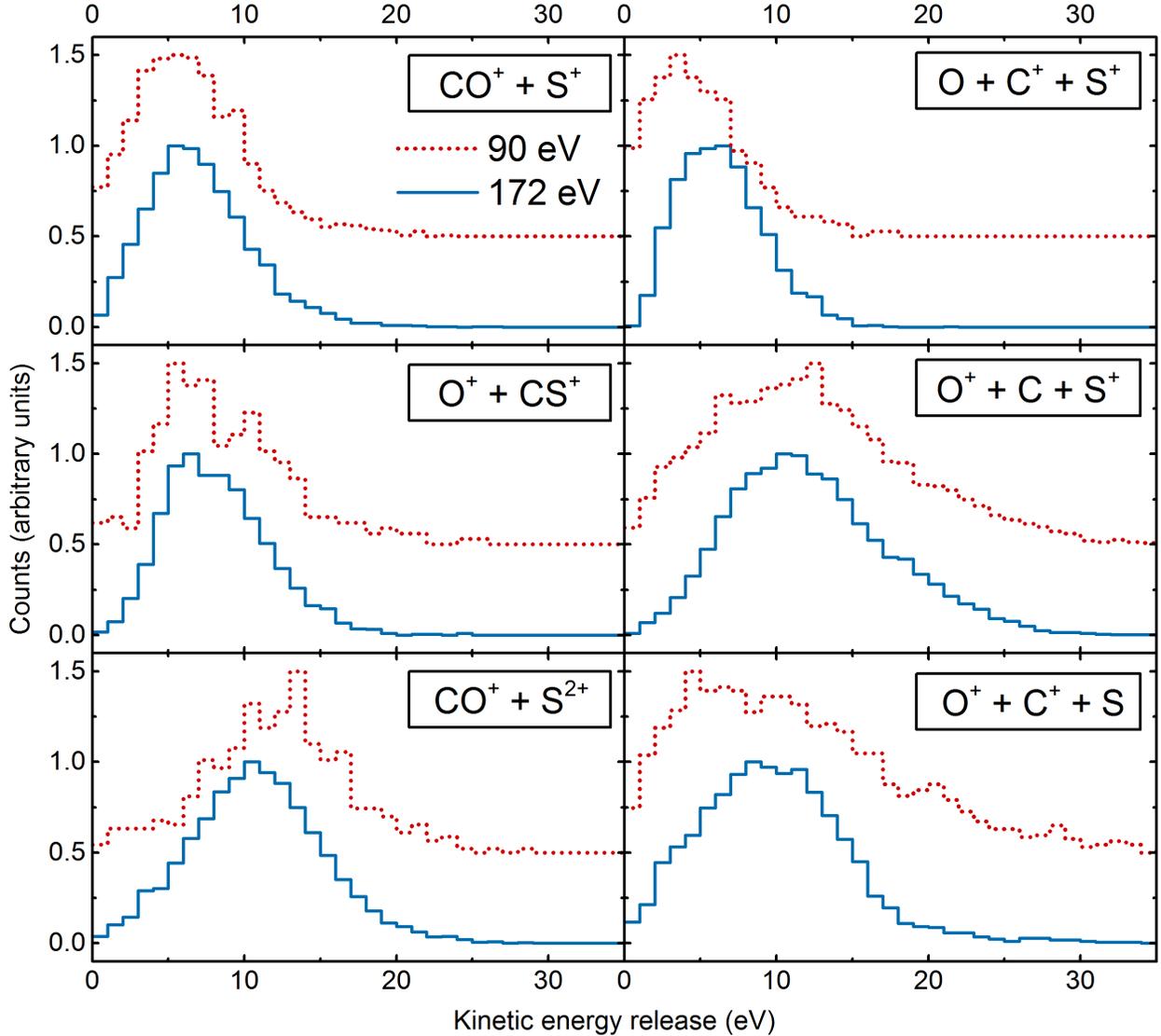

**Figure 1**. Kinetic energy release (KER) distributions of the three possible 2-ion three-body $OCS^{2+}$ fragmentation channels; (1,1,0), (1,0,1), and (0,1,1), and the KER distributions of the observed 2-ion two-body $OCS^{q+}$ ($q=2,3$) fragmentation channels; $O^+ + CS^+$, $CO^+ + S^+$, and $CO^+ + S^{2+}$. The dotted curves, corresponding to 90 eV excitation, are offset by +0.5 for clarity. Each distribution is individually normalized such that the maximum is 1.

For the 2-ion 3-body fragmentation channels that included a neutral atom, only the energy of the two ions is included in the total energy release plots, in order to compare with previous work [48,65] but the complete energy release can be calculated from conservation of momentum. In practice the difference is negligible with the neutral taking away less than 0.5 eV.

The events used in the creation of figure 1 all come from fragmentation processes in which only two ions are created, this is ensured by the requirement for conservation of momentum. In the case of the detection of two atomic ions, the third undetected fragment must be neutral. There cannot be any contamination by triple-ion fragmentation with an undetected third ion, as such an undetected ion would carry ~10 eV of kinetic energy with it and so the two detected ions by themselves could not conserve momentum.

|  | Current work (90 eV) | Current work (173 eV) | Saha et al. [40] (173 eV) | Masuoka et al. [56] (90 eV) |
| --- | --- | --- | --- | --- |
| $CO^+ + S^+$ | 7 eV | 7 eV | 6-7 eV | 5 eV |
| $O^+ + CS^+$ | 5, 10 eV (bimodal) | 7 eV | 7-9 eV | ~6 eV |
| $C^+ + S^+$ | 4 eV | 5 eV | 10 eV | 6, 10 eV (bimodal) |
| $O^+ + S^+$ | 10 eV (broad, asymmetric) | 10 eV (broad) | 10 eV | 7, 10, 20 eV (trimodal) |
| $O^+ + C^+$ | 5 eV (long tail) | 10 eV (broad) | 10 eV | Not observed |

**Table 1.** Qualitative comparisons between our measured KER distributions in figure 1 and those measured by Saha et al. [48] and Masuoka et al. [65] highlighting the peak KER energies and the qualitative nature of the observed distributions.

We observe several points of comparison between our KER distributions in figure 1 and those of Saha et al. [48] and Masuoka et al. [65]. Qualitative comparisons regarding the KER distributions are detailed in Table 1. We first note that excitation at 173 eV produces relatively simple KER distributions with a single peak. At 90 eV, multimodal distributions are observed indicating that several different electronic states are being accessed, in agreement with the observations of Masuoka et al. [65]. Magnuson et al. [41] have calculated potential energy curves for OCS indicating that the energy available to the molecule is greater than that found in the measured KER in figure 1. This suggests that these processes are not limited by energetics, but rather by dynamics. Furthermore, this also suggests that several electronic states are being accessed which all dissociate and auto-ionize at different rates. Overall, we find nothing to contradict earlier observations. Additionally, in the case of S(2p) or 172 eV excitation, the Auger decay seems to result in a narrower KER distribution, or at least a more symmetric distribution relative to the case of 90 eV excitation.

3.2 (1,1,1) kinetic energy release
The total KER by $OCS^{3+}$ fragmenting into three singly charged ion fragments is shown in figure 2 for multiple X-ray energies. The arrow marked "C" indicates the expected KER when fragmenting from the most probable ground-state geometry ($r_{CO}$=115.78 pm, $r_{CS}$=156.01 pm [66], and $\theta$=175° [3]) assuming a purely Coulombic potential and point-like particles [67]. For the (1,1,1) fragmentation channel of $OCS^{3+}$ we find this Coulombic KER to be 26.96 eV. The KER distributions for each photon energy peak at energies between 24 and 26 eV, making the (1,1,1) KER 89-96% of the Coulombic value.

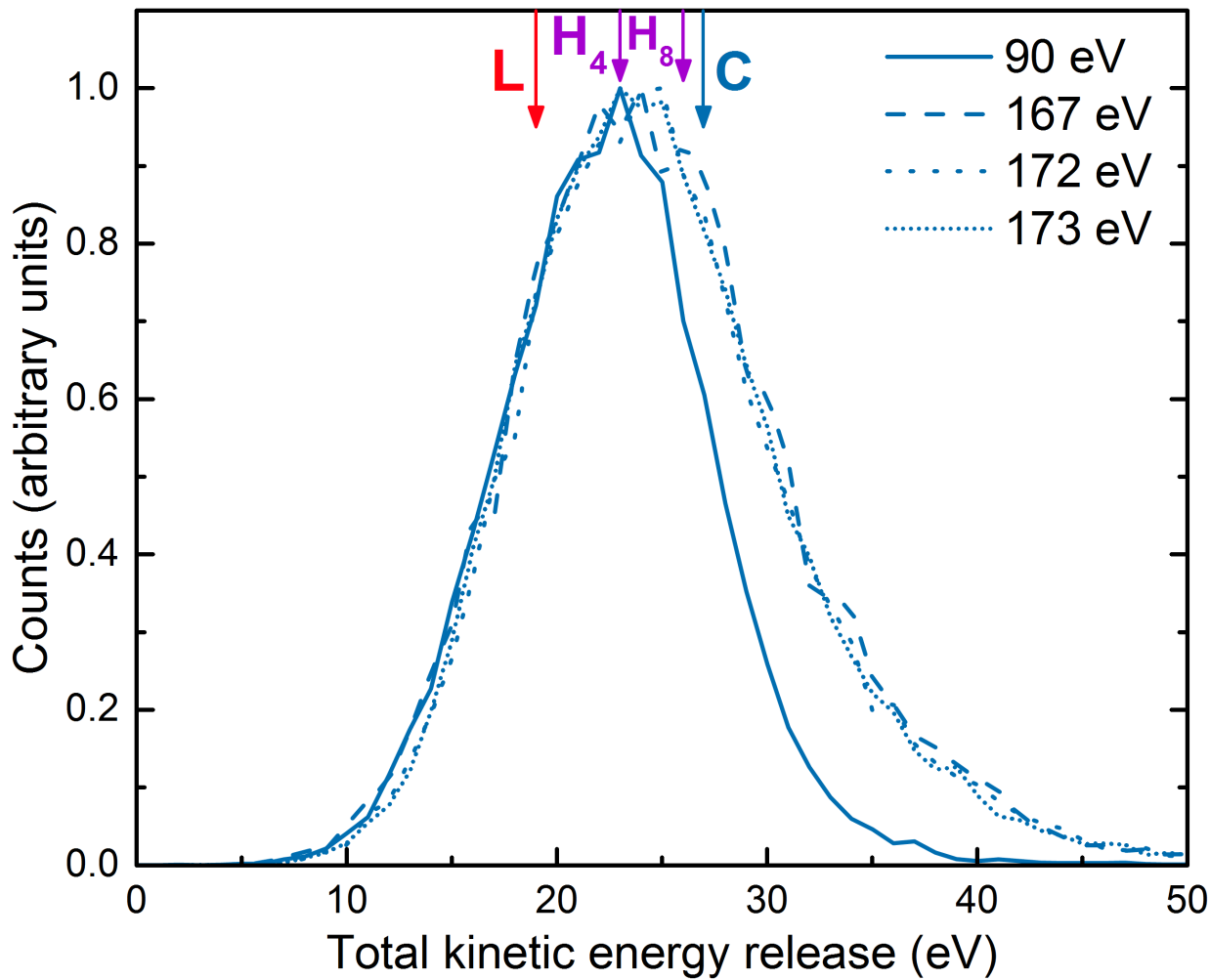

**Figure 2.** The total kinetic energy release (KER) distributions for the (1,1,1) fragmentation channel of $OCS^{3+}$ at 90, 167, 172, and 173 eV excitation. Arrow C indicates the expected KER (26.96 eV) when fragmenting from the ground-state geometry assuming a purely Coulombic potential and point-like particles. Arrow L indicates the experimentally measured KER after ionization by a few cycle infrared femtosecond laser pulse [59]. Arrows $H_1$ and $H_2$ indicate the experimentally measured KER for ionization by highly charged Argon ions, $Ar^{4+}$ for $H_4$ and $Ar^{8+}$ for $H_8$ [3].

The KER distributions for each excitation energy peak at approximately the same energy (24-26 eV), which is slightly less than that of a perfectly Coulombic explosion (indicated by arrow C). KER distributions from previous highly-charged ion (HCI) impact work peak at a similar energy (23 and 26 eV, indicated by arrows $B_1$ and $B_2$ respectively) to those in figure 2, with impact by $Ar^{8+}$ peaking at the higher energy (26 eV) and being more Coulombic than impact by $Ar^{4+}$ which peaks at 23 eV [3]. KER distributions for the (1,1,1) fragmentation channel of $OCS^{3+}$ by femtosecond induced Coulomb explosion [59] always peak at lower energies (maximum of 19 eV indicated by arrow L) than those of figure 2 even for very short pulses (7 fs).

From this we can identify that the X-ray and HCI impact excite a similar group of excited states but the femtosecond laser pulse experiment [59] accesses preferentially not only the ground state of the 3+ molecular ion, but its outer regions where the quasi bound nature is dominant [37,58]. Both exhibit a small tail although the KER distribution of $Ar^{4+}$ impact is quite symmetric while that of $Ar^{8+}$ is noticeably asymmetric about their peaks. They also exhibit a steeper low-energy rise and a shallower high-energy fall with a significant tail for all pulse lengths.

In comparing the overall energy distributions resulting from the different photon energies we immediately notice that the maximum KER for the (1,1,1) fragmentation channel from 90 eV excitation is 40 eV whereas for 167, 172, and 173 eV it goes up to 50 eV. It also has a steeper high-energy falloff than the KER distributions for other excitation energies, suggesting that energetics are limiting the available KER. This is a consequence of conservation of energy as the total amount of energy available to the molecule is determined solely by the absorbed photon from the X-ray beam.

In the case of excitation at 90 eV it takes 49.520 eV to produce $O^+ + C^+ + S^+$ from OCS with three zero energy electrons leaving a maximum of 40.48 eV available for the fragments and electrons. To obtain the value of 49.520 eV, we used the fact that the bond dissociation energy of S-CO to produce S + CO is 3.120 eV, and for C-O to produce C + O is 11.162 eV [68,69]. The first ionization energies of C, O, and S are 11.260 eV, 13.618 eV, and 10.360 eV respectively [70], giving a total of 49.520 eV. This corresponds with the triple ionization threshold measured by Masuoka [31] of 52.3 eV. Thus when excited by a 90 eV photon a maximum of 40.5 eV is available to the (1,1,1) KER. The KER measured for the (1,1,1) fragmentation channel at 90 eV cuts off at almost exactly 40 eV, supporting this argument. To our knowledge this is the first time that such a clear KER cut off has been observed, associated with a limit to the input energy.

Laksman et al. measured the total KER for excitation near the C(1s) edge [53]. Their KER distributions for $OCS^{3+}$ looks quite similar to our 167, 172, and 173 eV data, but with possible calibration differences. They show a peak from 10 eV to 30 eV with the high energy cut off having a slightly elongated tail compared to the low energy rise. Their KER distributions peak at a slightly lower energy (by only about 3-5 eV) compared to the current results in figure 2.

3.3 (1,1,1) geometry and dynamics
To identify the possible break up channels for the $OCS^{3+}$ states in more detail we use the Dalitz plot method, a well-established technique for displaying the possible molecular dissociation geometries in a two dimensional histogram [71]. Dalitz plots for the (1,1,1) channel are shown in figure 3. The experimental signal peaks near the red crosses at (0.11,-0.29) which indicate the location on the Dalitz plot of a purely Coulombic fragmentation, that is, when fragmenting from the ground-state geometry assuming a purely Coulombic potential and point-like particles. It is important to note that the geometry of the molecule and the momentum vectors are related in a nontrivial way and so the molecular bond angle is not the same as the angle between the momentum vectors of the terminal ions although sometimes simplified to such [53]. The vector angle is typically smaller than the molecular bond angle but can be larger for extreme bending.

Bending of the molecule in a concerted dissociation process, the most common process, results in a reduced angle between the outside momentum vectors and is associated with approximately vertical changes from the equilibrium point on the Dalitz plot.

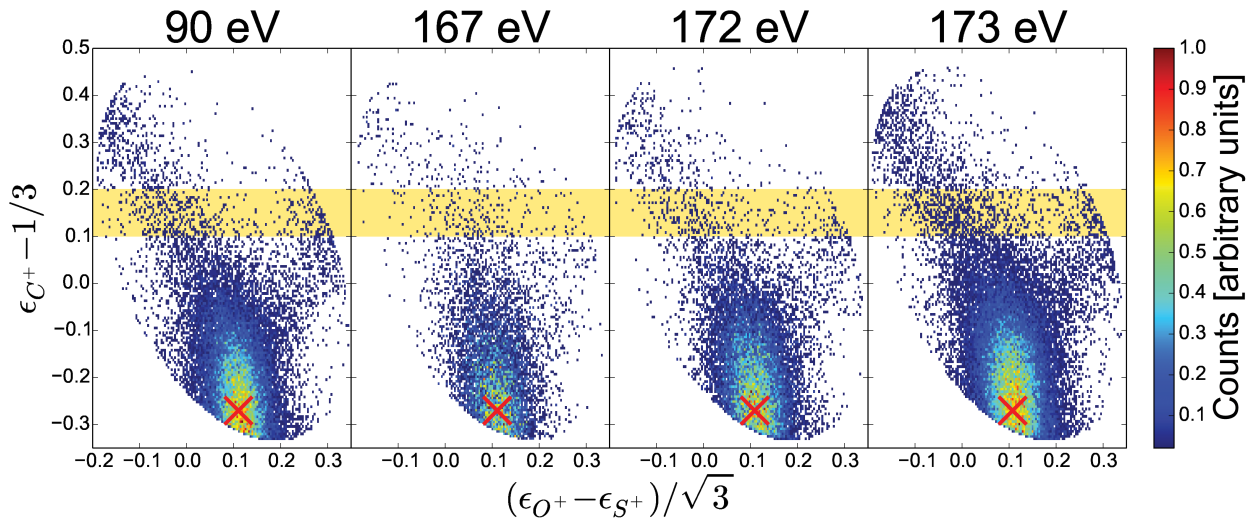

**Figure 3**. Dalitz plots for the (1,1,1) fragmentation channel of $OCS^{3+}$ at 90, 167, 172, and 173 eV excitation. The x-axis indicates the difference between the fraction of total energy released by the O and S ions divided by a scaling factor, and the y-axis is the fraction of energy released by the C ion minus an aesthetic factor to get the plot centered about *y*=0. Data was used exclusively from the region highlighted in yellow to produce a plot of the number of counts along the x-axis in figure 4. Each Dalitz plot is individually normalized such that the maximum is 1. The red crosses indicate the location on the Dalitz plot of a purely Coulombic fragmentation, that is, when fragmenting from the ground-state geometry assuming a purely Coulombic potential and point-like particles.

The significance of areas to the left or right of the equilibrium point have a more subtle origin, there is not a direct relationship to asymmetry of bond length but there is an association with the simultaneity with which bonds break, also known as the concertedness of the bond breaking reactions. As an illustration of this relationship, fragmentation of $OCS^{3+}$ into a metastable $CO^{2+}$ ion which itself breaks up after some life time (tens to hundreds of femtoseconds [72]), appears as a diagonal region, stretching from the bottom right (around 0.25,-0.33) to the top left (-0.15, 0.35). The neat linear relationship between how stepwise the bond breaking occurs and the amount of bending is simply related to conservation of angular momentum causing the metastable molecular ion to rotate as it separates from the $S^+$ ion. The amount of rotation experienced by the metastable ion before breakup correlates to different positions along the diagonal arms on the Dalitz plot, though not homogeneously. Therefore statistics are built-up with a higher density at the edges of the sequential features.

Three fragmentation channels are visible in the Dalitz plots for 90, 172, and 173 eV excitation; concerted breakup and two sequential breakup channels. The first sequential process proceeds via the metastable $CO^{2+}$ ion, as described above and the second via a metastable $CS^{2+}$ molecule, which appears as a diagonal region from the bottom left towards the top right. It is worth noting that both of the stepwise channels are strong in the case of $OCS^{3+}$ and are represented on a linear intensity scale, whereas for the $CO_2$ and $CS_2$ molecules [72,73] the stepwise process required the use of a logarithmic scale. Sequential breakup channels appear weak in the Dalitz plots of Laksman et al. [53] for the case of C(1s) excitation and for highly charged ion impact work [3] although this may be due to the method of trigger which can lead to loss of sensitivity to the generally low energy process. Both of these cases may be resolved by further experimental attention.

In the current work, the 167 eV data shows a much lower stepwise signal which we will return to later. By comparison, Dalitz plots produced from previous femtosecond laser work [59] show narrower and more pronounced arms suggesting that the metastable molecular ion is more vibrationally excited by the synchrotron ionization processes, leading to increased range of energy release by its breakup.

The final dissociation process, the concerted breakup in which both bonds break at the same time, is the most significant channel. In the Dalitz plots of figure 3 the concerted process is represented by a zone from the bottom edge up to about zero on the y axis. Although as mentioned above, the peak signal is found to be near to the equilibrium geometry, the vertical and horizontal spread are consistent with bending and a degree of stepwise breakup respectively. This is a similar region to that exhibited under ionization by 7 fs laser pulses and HCI impact [3,59] and in contrast to the progressively vertically stretched region characteristic of longer pulses (up to 200 fs) where the molecule has time to bend in the laser field, and indeed is bent by the field [59].

The concerted process offers us the best opportunity to see differences due to the slower (10 fs) multiple ionization event associated with the Auger process, which therefore affords increased time between ionization steps and so more time is spent on the singly and doubly ionized potentials than is possible for the direct ionization process (<1 fs). The result of these increased residence times would include a greater degree of stepwise nature and increased bending exhibited in the Dalitz plots for 172 and 173 eV compared to the direct process best exemplified by 90 eV. There are indeed small but discernable differences between the direct and Auger ionization plots, particularly in the vertical direction, but these differences can be seen more clearly by integrating over the vertical axes and constructing a new plot shown in figure 4.

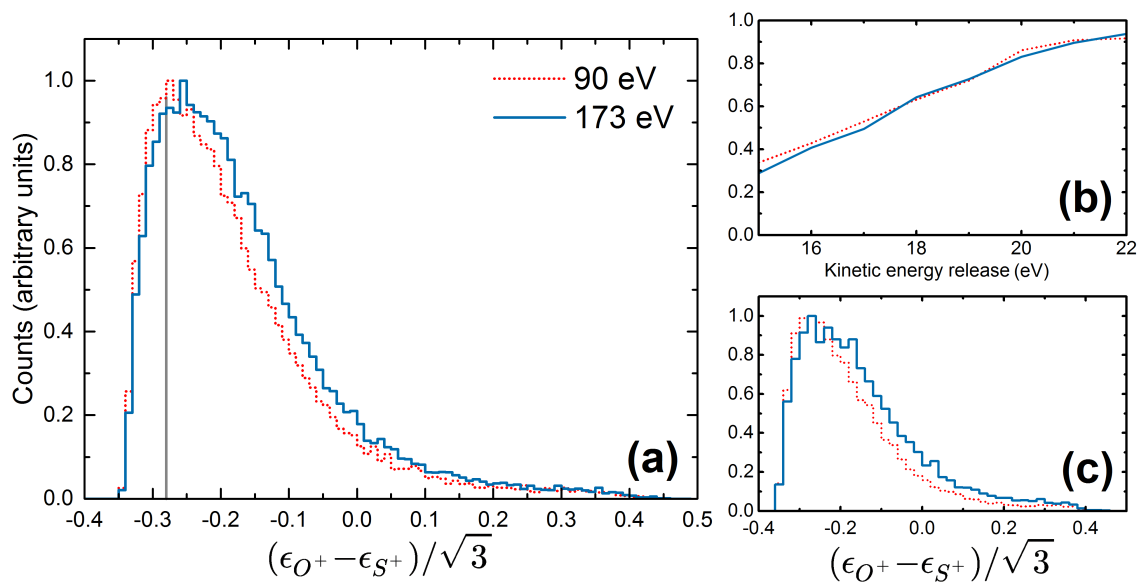

**Figure 4**. (a) Projections on the y axes of the Dalitz plots of figure 3 comparing the direct ionization process of 90 eV with the Auger ionization process at 173 eV. Each curve is individually normalized such that the maximum is 1. Gray drop lines indicates the location on the Dalitz plot of a purely Coulombic event, from equilibrium geometry. (b) Region of figure 2 for which the kinetic energy release from 90 eV and 173 eV is nearly identical. (c) Same as (a) but from data limited to the region of identical kinetic energy release.

As expected, the 173 eV data shows both more bending in the form of a peak which is at higher Y value and a wider distribution than the 90 eV data. The shift of the peak although small is significant as the 90 eV distribution is peaked very close to the predicted equilibrium value, marked by the gray dropline, indicating that for the fastest multiple ionization events at 90 eV, very little motion is possible for the molecule during the ionization process whereas for 173 eV even the fastest multiple ionization event allows time for some bending to occur. The conclusion that this is a time-scale effect depends on the assumption that the states of the molecular ions occupied are the same. The only certain way to determine this is to observe the electrons emitted during the ionization process, but in the absence of this a good indicator is the total kinetic energy released by the ionic fragments. Although the patterns of energy release are very similar for all photon energies, we can restrict the energy range over which we compare the bending parameter to one where the total kinetic energy release for 90 eV and 173 eV photons is nearly identical. This region is shown in figures (4b) and (4c) shows the corresponding comparison between the Dalitz plot Y projection exhibiting the same enhanced peak bend and wider range of bending.

The data at 167 eV which is just below the S(2p) edge has some significant differences to the other plots in figure 3. For 167 eV the stepwise arms appear diminished and a single central distribution stretching upwards is apparent. This difference becomes clear when we consider the highlighted region in figure 3, for which the signal is plotted in figure 5 integrated in the vertical direction. At a photon energy of 167 eV, the distribution is dominated by a maximum peaking close to $x=0.1$ compared to the other energies which are dominated by two maxima near $x=0$ and $x=0.3$, each corresponding to an arm on the Dalitz plot. Although some stepwise events still exist at $x=0.3$ for 167 eV, the dominant distribution at $x=0.1$ represents processes leading to enhanced bending.

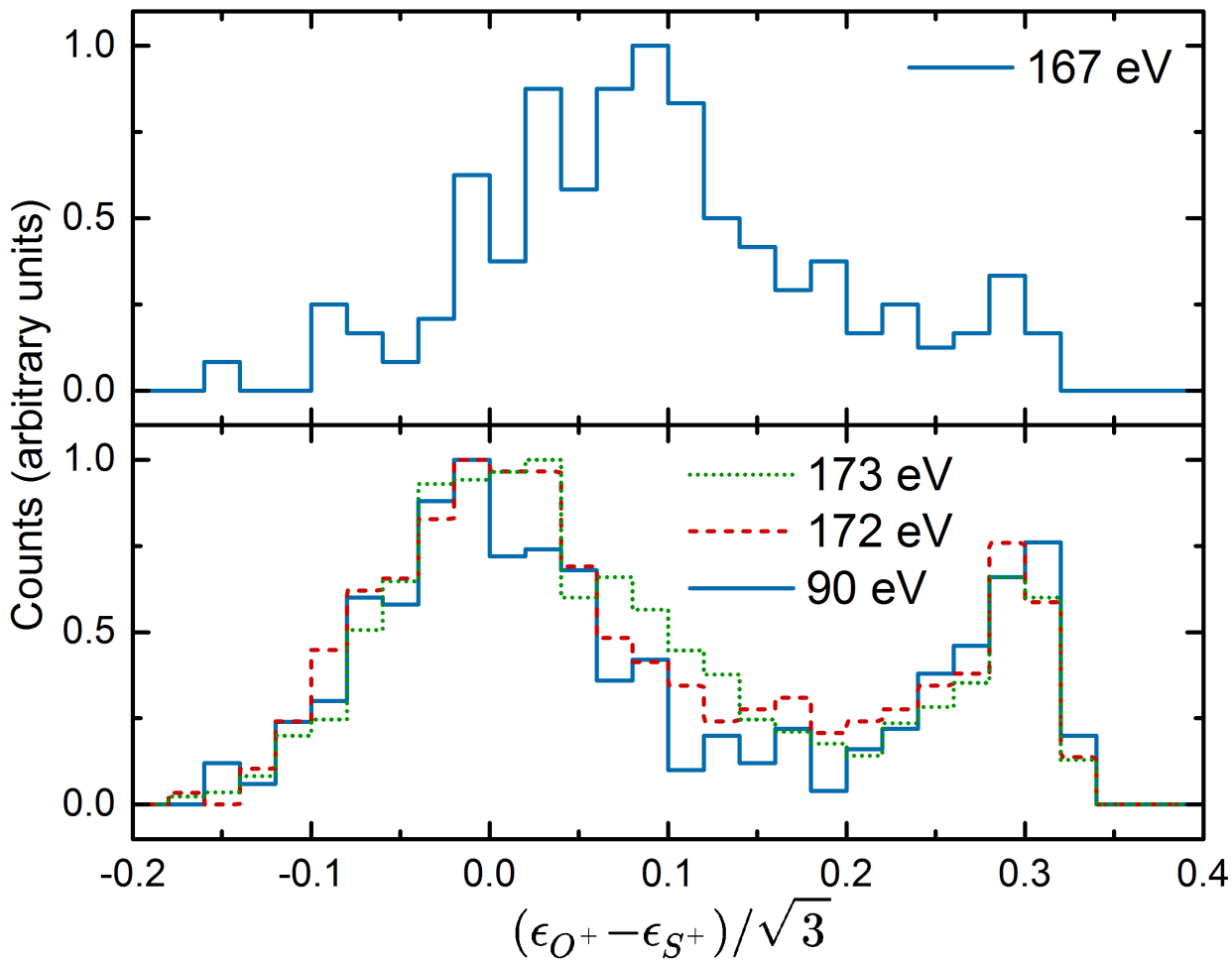

**Figure 5**. A count of the number of events with $0.1 \leq \epsilon_{C+} - ⅓ \leq 0.2$ from the region highlighted in yellow in figure 3 for photon energies of 90, 167, 172, and 173 eV. The two arms due to sequential breakup (or stepwise) processes in the 90, 172, and 173 eV Dalitz plots can be seen as two peaks at $x=0$ and $x=0.3$. However, for 167 eV only a single peak can be observed at $x=0.1$.

In order to gain some insight into the difference, we consider that excitation by a 167 eV photon corresponds to exciting OCS to the $(2p_{3/2})^{-1} \rightarrow 4s$ Rydberg state [40]. The Rydberg state (and the direct ionization) should only show vibrational excitation in the symmetric stretch (v1) and asymmetric stretch (v3) modes. The molecular state we are exciting to is linear and so we do not expect to observe any bending dynamics. Therefore as a major channel we should see no difference between 90, 167, 172, and 173 eV.

However the OCS molecule may be excited by a 167 eV photon to other states with similar energies. Excitation at $(2p_{3/2})^{-1} \rightarrow 4s$ (167 eV photon) sits on the tail of the highly bent $\pi^*_{1/2}$ state at 165.4 eV. Thus as well as exciting the 4s Rydberg state which is expected to be linear, we are additionally exciting the $\pi^*_{1/2}$ state which is expected to be highly bent due to the strong interactions between the excited electrons and the remaining core electrons [40]. Thus following excitation at 167 eV we see the stepwise arms due to the excitation of the 4s Rydberg state, and we additionally see a central distribution stretching upwards due to excitation to the $\pi^*_{1/2}$ state. This central arm is indicative of a bending process occurring. Typically, bending of molecules during single photon ionization is accepted to occur but be incomplete on the timescale of the multiple ionization event, differing degrees of bending have been inferred using the photo electron-photo ion-photo ion-coincidence method (PEPIPICO) depending on the photon energy [74,75], though not because of differing timescales, but in the context of our current discussion, timescales are important. The extent to which the molecule is allowed to bend significantly during the fast multiple ionization process needs to be addressed, and with the current time and position sensitive method we can determine the full bend distribution. The important factor is that the peak in the Dalitz plot Y projection in figure 4(a) is still close to the equilibrium indicating that the fastest ionization events occur before any bending can happen, but the slower ionization events can image the molecular dynamics induced by the $(2p_{3/2})^{-1} \to 4s$ excitation. Slower multiple ionization events do of course happen, when metastable excited states are accessed, as with the stepwise processes, which can take hundreds of femtoseconds to complete.

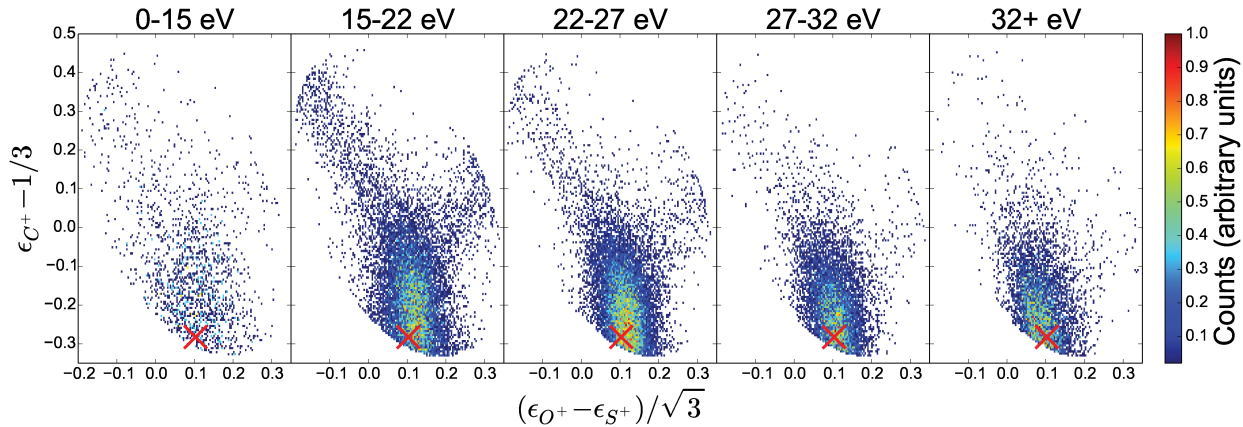

**Figure 6**. Dalitz plots for the (1,1,1) fragmentation channel of $OCS^{3+}$ at an excitation energy of 173 eV for different ranges of KER. The red crosses indicate the location on the Dalitz plot of a purely Coulombic event, that is, when fragmenting from the ground-state geometry assuming a purely Coulombic potential and point-like particles.

In order to consider the connection between the KER and the dynamics taking place during breakup, figure 6 shows the Dalitz plot for 173 eV data sliced into five energy ranges. At low values of KER (0-15 eV), the Dalitz distribution is almost uniformly distributed around the simulated Coulombic center indicated by a red cross. There are events in the regions associated with stepwise processes but the arms are not well-defined. At mid values of KER (15-22 and 22-27 eV), the concerted distribution is centered a little to the right of the Coulombic simulation ($x$=0.11) in contrast to the overall distribution which is centred horizontally close to this equilibrium point as indicated by figures 3 and 4. This implies that for molecules near equilibrium in the bending coordinate, the S-C bond is breaking shortly before the CO bond resulting in relatively more energy deposited in the $O^+$ ion then the $S^+$ ion. The long timescale stepwise arms are well defined and strongest in the 15-22 eV range showing that they originate from excitation to specific metastable states which can only result in significantly lower than Coulombic KER. A possible mechanism at this KER range is that during the Auger cascade a dication is formed as $OCS^{2+}$ rapidly dissociates to $S^+$ + $(CO^+)^*$ which at a later stage during the dissociation process autoionizes to $CO^{2+}$ to eventually form $C^+$ + $O^+$. It is interesting to note that the stepwise arms are most prominent in the 15 eV to 22 eV KER range, which overlaps with, or is only slightly greater than, the 2 ion KER (section 3.1). At higher values of KER (27-32 and 32+ eV) the concerted process shifts to the left until it is centered horizontally on the simulation, indicating a higher degree of simultaneity, consistent with the assumption of instantaneous bond breaking.

This is the first such observation of a horizontal shift in the position of the concerted process but inspection of [50] shows the existence of an overall shift for the few cycle laser results which also exhibit overall lower peak in KER consistent with the trend observed here. A further significant feature, of the plots is the progressive restriction of the concerted region in the vertical direction, towards the equilibrium point, with increasing KER, which is consistent with both bonds breaking more rapidly.

3.4 $OCS^{4+}$ kinetic energy release

The KER distributions for the three $OCS^{4+}$ channels producing triple ion fragments are shown in figure 7 for 173 eV excitation.

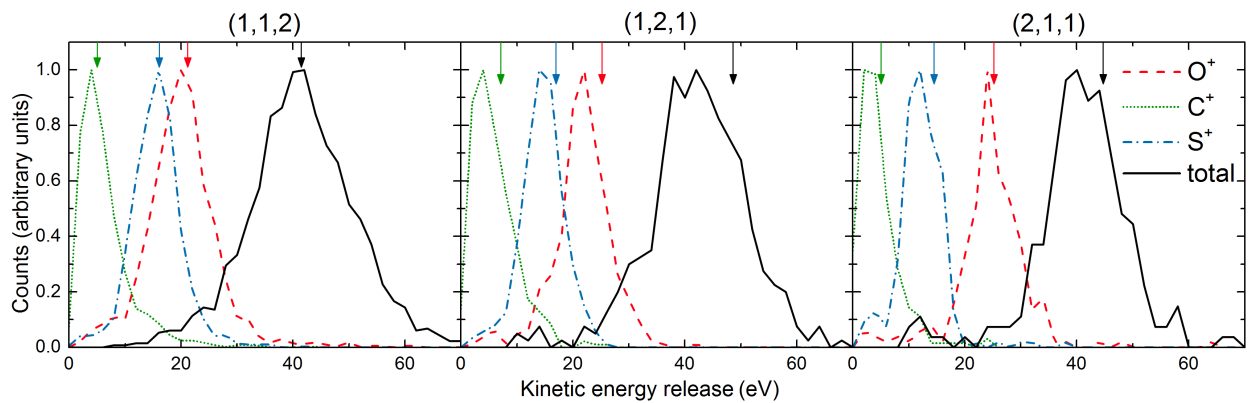

**Figure 7**. KER distributions for the three $OCS^{4+}$ channels producing triple ion fragments; (1,1,2), (1,2,1), and (2,1,1) at 173 eV excitation. The total KER distributions are plotted as well as the KER distributions of the individual fragments. The arrows indicate the expected KER when fragmenting from the ground-state geometry assuming a purely Coulombic potential and point-like particles.

$OCS^{4+}$ events are significantly rarer than $OCS^{3+}$ events (10-100 times rarer) hence the poorer statistics. The most commonly observed fragmentation channel of $OCS^{4+}$ is the (1,1,2) channel, followed by the (1,2,1) and then the (2,1,1) channel. This can be explained by the increasing second ionization energies of S, C, and O, which are 23.34, 24.38, and 35.12 eV respectively [70]. These distributions are also very similar at 167 and 172 eV. At 90 eV excitation, $OCS^{4+}$ production is energetically unfavourable and so events are extremely rare and the fragment ions are left with little kinetic energy. The KER distributions in figure 7 all have similar shapes and peak at similar energies (40-42 eV). They are broader than the KER distributions for the (1,1,1) fragmentation channel in figure 6. The (1,1,2) distribution peaks exactly at the Coulombic value, while the (2,1,1) distribution peaks at a slightly lower energy than the Coulombic value, and the (1,2,1) distribution peaks at a noticeably lower energy than the Coulombic value.

3.5 $OCS^{4+}$ geometry and dynamics

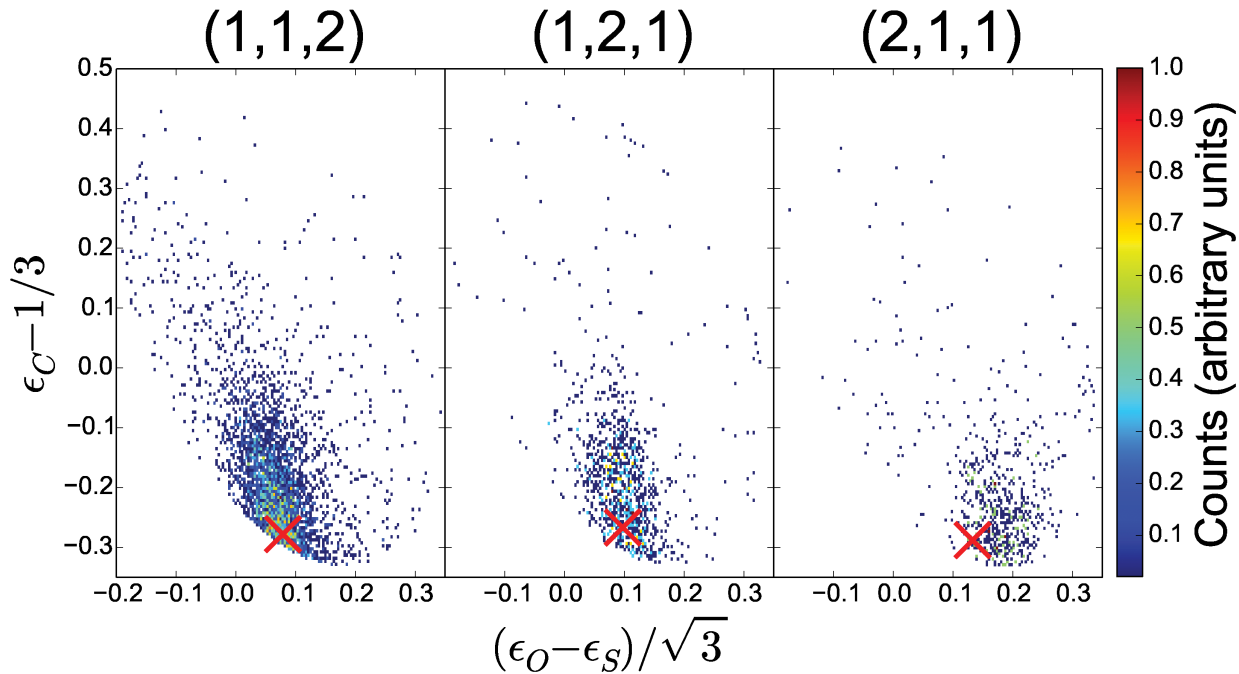

**Figure 8.** Dalitz plots for three possible fragmentation channels of OCS$^{4+}$ at a photon energy of 173 eV. The red crosses indicate the location on the Dalitz plot of a purely Coulombic fragmentation, that is, when fragmenting from the ground-state geometry assuming a purely Coulombic potential and point-like particles.

Dalitz plots for the three possible fragmentation channels of OCS$^{4+}$ at an excitation energy of 173 eV are plotted in figure 8. In the case of the (1,1,2) fragmentation channel, the majority of observed events are concerted and centred around the Coulombic simulation indicated by the red cross, some CO$^{2+}$ metastable sequential events exist giving a diagonal background. The (1,2,1) fragmentation channel is similarly centred around the equilibrium position and no stepwise channels are possible. The (2,1,1) fragmentation is unusual however because it is far from centred on the equilibrium cross and shifted to the right which means the O$^{2+}$ ion carries a significantly higher fraction of the total KER than for the equilibrium Coulombic case. As in the less extreme case of the low energy release (1,1,1) channel this can be seen as indication that the CS bond breaks before the CO bond. In order to formulate a possible cause for this phenomenon, we should first consider how an initial ionization process which results in removal of a core 2p electron from the sulphur atom can finally result in fragmentation involving a more highly ionized oxygen ion than the sulphur ion, despite its distance from the initial ionization site and its higher double ionization potential. In order to do this we should consider the molecular orbitals of neutral OCS [76]. The HOMO (2π) is anti-bonding with respect to the OC bond and has considerable overlap with the S. Thus involvement (loss) of this valence orbital in the Auger cascade will result in relatively strengthening and shortening the O-C bond. The HOMO-2 (1π) is extremely bonding with respect to O-C and has negligible overlap with the S, thus the retention of this orbital during the Auger cascade will again result in a shortening of the O-C bond and strengthening it with respect to the CS bond, while increasing the amount of charge being retained near the O atom. As mentioned earlier, the Dalitz asymmetry cannot be simply linked to asymmetry of bond length, but implies breaking of the CS before the CO bond as a result of the asymmetry inducing cascade. To our knowledge this is the first such observation.

## 4. Conclusions

We have attempted to compare the ionization by direct (90 and 167 eV) and Auger process (172 and 173 eV) resulting from the S(2p) core ionization, in order to discern any timescale effects which might be manifested in the plethora of processes which are initiated. Both methods can be thought of as ultrafast in terms of the speed with which electrons are removed from the molecule, but the direct process has the potential to be the faster, at the femtosecond scale as opposed to the tens of femtoseconds. For the low ionization processes we see little difference between the current measurements and those of previous work [31,48] irrespective of the photon energy, although we are able to add considerable detail due to the high photon flux available in the CLS. The consistency is perhaps not surprising given that these processes release relatively little energy as they are not dominated by Coulombic potentials and are therefore relatively robust to time scale changes. Even for triply ionized states the major difference between energy release is dominated by the limitation imposed by a photon energy of 90 eV. It is only when we look at the pattern of momentum release in comparing 90 eV and 173 eV Dalitz plots that we can see distinct evidence of timescales effects. A pattern, particularly imprinted in the peak position of the bending distribution, indicates that there is very little dynamics before ionization is complete, for the direct process, but that there is a small but significant bend characteristic of stepwise ionization, present for the Auger process. In fact the equilibrium bond angle of OCS is remarkably well imaged for the (1,1,1) channel by the peak of the energy integrated Dalitz plot, this is in contrast to the results of 7 fs laser pulses [59], suggesting that the ionization process is considerably faster for the synchrotron case, in fact this would be faster than any tabletop multiphoton systems [77] which show great promise in imaging molecular dynamics on femtosecond time scales. Our results show that below the S(2p) edge, we can see ionization events which involve metastable steps and result in large amplitude angular motion in either concerted (167 eV) or stepwise processes. We have identified, for the first time, a connection between low KER and breaking of the SC bond before the CO bond. Finally for the 4+ ionization, we see evidence that in generating the (2,1,1) channel the molecule is excited to even more asymmetric dissociation. This phenomenon is only observable because of the large photon flux, but indicates that more exotic phenomena may be observable using other energies available at the third generation synchrotron source. In conclusion we have been able to identify a surprisingly rich set of dynamics, on a femtosecond timescale resulting from Auger decay after ionization of the S(2p) electron and see evidence that direct ionization takes place on shorter timescales. In order to confirm the exact timescales, with which the processes proceed, it is possible that pump probe studies could be carried out utilising pulses from a free electron laser or high harmonic attosecond source.S

**Acknowledgments**

We gratefully acknowledge research funding from NSERC (Natural Sciences and Engineering Research Council of Canada). Part of the research described here was conducted at the CLS, which is supported by the Canadian Foundation for Innovation, the Natural Sciences and Engineering Research Council of Canada, the University of Saskatchewan, the Government of Saskatchewan, Western Economic Diversification Canada, the National Research Council Canada, and the Canadian Institutes of Health Research.